\documentclass[11pt,a4paper]{article}

\usepackage[T1]{fontenc}
\usepackage{amsmath,amssymb}
\usepackage{graphicx}
\usepackage{xcolor}
\usepackage{xurl}
\usepackage[round,authoryear]{natbib}
\usepackage{authblk}
\usepackage[colorlinks=true,linkcolor=blue,urlcolor=blue,citecolor=blue]{hyperref}
\usepackage{bookmark}

\title{Adolescents \& Anthropomorphic AI: Rethinking Design for Wellbeing\\
\large An Evidence-Informed Synthesis for Youth Wellbeing and Safety}

\author{Mathilde Neugnot-Cerioli\textsuperscript{\dag}}

\affil{\textsuperscript{\dag}Published under the name Mathilde Neugnot-Cerioli.
Also known professionally as Mathilde Cerioli.
Corresponding author: \texttt{mathilde@everyone.ai}}

\date{}

\begin{document}

\maketitle

\begin{abstract}
Conversational AI has become part of adolescents' everyday lives. This report asks: what does AI owe adolescents when it can speak to them like a social partner? The synthesis bridges the gap between developmental science and industry practice through consultations, a behavioral framework, and global policy dialogue. It identifies non-negotiable guardrails and highlights the role of anthropomorphism as a design lever for risk mitigation, ensuring systems support adolescents' autonomy and skill development.
\end{abstract}

\section*{Contributors}

This report benefited from insights, feedback, and discussions with the following contributors.

\noindent Adrien Ab\'ecassis, Chief Policy Officer, Paris Peace Forum; Kate Blocker, Ph.D., Director of Research and Programs, Children \& Screens; Maxime Derian, Ph.D., Research Associate, (C\textsuperscript{2}DH), Universit\'e du Luxembourg; Sara M. Grimes, Ph.D., Professor, Department of Communication Studies, McGill University; Thao Ha, Ph.D., Director, HEART Lab; Associate Professor of Psychology, Arizona State University; Sameer Hinduja, Ph.D., Co-Director, Cyberbullying Research Center; Professor, Florida Atlantic University; Dan Hipp, Ph.D., Senior Research Coordinator, Children \& Screens; Melinda Karth, Ph.D., Project Coordinator, Children \& Screens; Pilyoung Kim, Professor of Psychology, University of Denver; Maxime Le Bourgeois, Cognitive Neuroscience Researcher, Everyone.AI; Sonia Livingstone, Professor, Department of Media and Communications, London School of Economics; Polina Lulu, Child Experienced Researcher, Youth Voices; C\'eline Malvoisin, Head of Operation \& Programs, Everyone.AI; Olga Muss Laurenty, Researcher, Everyone.AI; Kris Perry, Executive Director, Children \& Screens; Gregory Renard, Board President, Everyone.AI; Bethany Robertson, Senior Advisor, Everyone.AI; Anne-Sophie Seret, Executive Director, Everyone.AI; Sonia Tiwari, Children's Media Researcher; Scott Traylor, Ed.M., Educational Researcher; Ying Xu, Assistant Professor of Education, Harvard University.

\section{Introduction and purpose}\label{introduction-and-purpose}

Conversational AI has turned language into an interface for everyday
life. In a single thread, a young person can get homework help, rehearse
what to say to a friend, ask for advice they would not voice aloud, or
look for comfort when they feel alone. That versatility opens up
opportunities when it comes to access personalized information, and it
also introduces risk. This new generation of tools are reproducing the
way humans are interacting and communicating, creating potential
confusion.

For adolescents, that shift lands in a sensitive window of cerebral
development. We define adolescents here as individuals aged 13 to 18
years. This range reflects the age at which most digital platforms
permit independent use and the period during which individuals remain
covered, with specific protections and obligations still applying up to
age 18. This developmental period is critical to identity formation,
when peer belonging and status become central, the drive for autonomy
develops faster than critical judgment, and emotional regulation is
still under construction \citep{C1K0}. In this transitional period,
supportive tools for adolescents represent a real opportunity, and even
more so for teens who are more socially anxious, isolated, or lack
access to trusted adults. But this also raises the stakes when AI
systems behave in ways that simulate intimacy, deliver
consequence-free validation, or drift into ``always there''
companionship dynamics.

This work asks a direct question: what does AI owe adolescents when it
can speak to them like a social partner? The report focuses on
adolescents and anthropomorphic conversational AI because that is where
the governance problem is becoming both concrete and urgent. Teens are
using general-purpose chatbots and companion-style systems at scale
\citep{VAom, le5W, zUm1}. The pattern of use is evolving quickly, and
more emotionally engaged interactions are becoming more common.
Meanwhile, the empirical literature on short- and long-term impacts is
still catching up to product cycles and adoption curves.

Given that gap, the goal of this document is practical. It translates
what we know, what experts converge on, and what remains uncertain into
a developmentally grounded framing for policy, design, and investment.
It builds on our earlier foundational work on child development in the
AI era \citep{Rj5w}, and it narrows in on what has become urgent for
adolescent-facing systems: the specific interaction cues that shape
trust, reliance, and the direction of a teen's social learning over
time.

The synthesis is built from the results of three complementary actions
impulsed and driven by our team over the course of the last six months,
each designed and chosen for what they can add at this stage built from
three inputs:

\begin{itemize}
\item \textbf{Consultations} to identify the mechanisms and design levers
  experts and industry teams and young people already see as high-stakes
  (persona, validation style, reminders, refusal and deflection,
  escalation on sensitive topics, engagement depth, and memory or
  personalization).
\item \textbf{The iRAISE Lab,} a two day in person workshop that convened
  industry and child experts, to turn those concerns into an assessable
  behavioral lens, using realistic teen scenarios to test how model
  responses shift along a gradient from tool-like support to
  relationship-like dynamics.
\item \textbf{A global multi-stakeholder dialogue at the 8th edition of the
  Paris Peace Forum} (an international diplomacy event held in October
  in Paris), to better understand how to develop the framing for real
  governance constraints across jurisdictions, sectors, and child-rights
  commitments.
\end{itemize}

A core thread runs through all three streams: adolescents will relate to
these systems socially, whether developers intend it or not. The
question is whether those parasocial dynamics, meaning this interaction
that feels social but is devoid of reciprocity, are being shaped toward
developmentally supportive use, or toward reliance patterns that
displace real relationships and weaken the very skills adolescence is
meant to build. This is why this report focuses on model behavior and
interaction patterns, rather than product types. A ``general chatbot''
can become a companion in private: a ``teen tool'' can drift into
intimacy if its defaults reward frequency of use, agreement, and
attachment.

\section{Methodological approach: Multi-stream evidence synthesis
and stakeholder elicitation}\label{methodological-approach}

This synthesis was built through an iterative, multi-stream process
designed to bridge two practical tensions. The first is the difference
in timelines between industry and research: AI systems are being
deployed and adopted faster than developmental science can produce
long-term evidence. The second is translation: the concepts that matter
in adolescence research are not simple rules that can be easily applied
to product decisions, where teams need quantifiable metrics and
clear-cut boundaries.

For that reason, the work began with the questions surfacing inside
product and safety teams, then used expert consultation,
multi-stakeholder workshops, and policy dialogue to test, refine, and
structure the answers to those questions into a coherent set of
mechanisms and governance levers.

\subsection{Industry Scoping: How relational should AI systems feel}\label{industry-scoping}

The first step was direct engagement with AI companies and product teams
to identify which aspects of youth--AI interactions felt most urgent,
consequential, and currently difficult to operationalize. This grounded
the work in the decisions teams are already making, rather than in
hypothetical risk inventories, and ensured that it first empowered the
actors who can implement change.

Across these conversations, the same set of practical questions came up
repeatedly: how ``human'' or relational an AI persona should be; when
and how AI should refuse, deflect, or ``punt'' on requests; whether time
limits or turn-taking limits are appropriate and how they should
function; how often AI systems should remind users that they are not a
person; when a model should provide direct answers versus scaffold the
user toward their own reasoning; and how to handle high-stakes
disclosures, particularly suicidal ideation. These were often raised as
feature-level decisions or edge-case safety problems. The population
identified as the most important to address first was adolescents,
specifically those above 13, as they are the youngest active users and
because 13 is often the legal age to use most of these models without
supervision.

Taken together, most of these issues converged on a larger question:
where the line sits between legitimate support and undue emotional or
relational influence. Many of the ``feature'' questions were actually
the result of parasocial dynamics and emotional over-reliance, and of
how anthropomorphic interaction patterns can change the felt meaning of
an AI system over time.

\subsection{Expert Convergence: Development, Mental Health, Rights, and Safety}\label{expert-convergence}

Following the industry-led scoping, the synthesis was anchored by
in-depth expert consultations (see Appendix~\ref{appendix-2} for the
list of experts). Experts spanned adolescent and adolescent relationship
development and neuroscience, child and adolescent mental health,
learning sciences and education, child-rights-based policy, online
safety and trust and safety practice, and AI governance. These
consultations do not replace empirical evidence, but by aggregating
relevant expertise, they have proven to provide a grounded and efficient
pathway to catching up with fast-moving product realities and informing
product and policy decisions through child-development-informed input.

A first key lesson was convergence across disciplines on what matters
most developmentally. Many risks raised by industry were echoed by
developmental and mental health researchers, especially concerns around
persona design, unconditional validation, prolonged engagement, and
emotionally intimate interaction patterns that are not tightly bound to
a specific use or purpose. Experts repeatedly emphasized that these
design features are not just UX preferences, and that they intersect
directly with core developmental tasks by shaping how users interact
with AI systems.

This also reframed what ``wellbeing'' means in this context. Rather than
treating wellbeing as a narrow clinical endpoint, experts emphasized
wellbeing as capacity: an adolescent's ability to navigate the central
work of this life stage, including learning social norms through real
and sometimes raw feedback, forming identity in relation to peers,
developing autonomy and judgment, tolerating frustration and discomfort,
and cultivating independent thought. From this perspective, the
highest-impact design risks are those that systematically remove social
friction, simulate intimacy, or deliver consequence-free validation in a
way that can weaken resilience and independent thinking over time.

That framing clarified how to interpret the feature questions that
started the process. Time limits, turn-taking, reminders, response
style, and crisis escalation shall not be regarded just as isolated
safety mechanisms: we will consider them as the quantifiable and
interdependent levers that can either keep the system in a bounded,
tool-like role or drift it toward relationship substitution, with very
different developmental implications.

\subsection{From shared concerns to a structured research agenda}\label{shared-concerns}

By starting with industry questions and then reframing them through
expert input, the process aimed to avoid a potential issue common in
safety work: treating symptoms while missing the underlying dynamics.
The iterative dialogue made clear that many surface questions cannot be
answered well without a coherent theory of how adolescents relate to
these systems over time, particularly when the interaction becomes
emotionally engaged.

This directly shaped the design lab work and the construction of the
shared framework. A pragmatic rationale was agreed upon: tackle
high-leverage design choices that teams can adjust, and which strongly
influence whether adolescents experience the system as a bounded tool or
as something closer to a social partner. The overall goal was defined as
to move from general claims to operational variables by centering the
analysis on anthropomorphic cues, relational cues, and engagement
dynamics.

Emotional reliance and parasocial interaction were purposefully treated
as the first analytical layer in the synthesis, since they function as a
web linking otherwise disparate design decisions, and mark the point
where both industry and researchers independently identified the highest
potential stakes for adolescent wellbeing.

\subsection{A global multi-stakeholder dialogue to test the agenda in the real world}\label{global-dialogue}

Subsequently, a high-level multi-stakeholder dialogue convened at the
Paris Peace Forum in October 2025 was used to evaluate how to best
translate current efforts into the global governance setting. The
session brought together participants from governments (France),
companies (OpenAI, Google, Orange), research organizations, NGOs
(Joan-Ganz Cooney Sesame Center at Sesame Workshop, Human Technology
Foundation), international institutions (UNICEF, UNICRI), and Youth
Voices, with the explicit goal of assessing whether the framing holds
when examined across jurisdictions and governance traditions (see
Appendix~\ref{appendix-1} for complete list of participants and full
recordings).

The dialogue was structured around three elements of the agenda: the
children's-rights anchor, the role of anthropomorphism as a risk
pathway, and the focus on model behaviors and interaction patterns
(rather than relying primarily on content categories). Participants were
invited to discuss how this model-behavior-modification framing holds
across different legal and cultural contexts, identify where the framing
might require adaptation to remain meaningful in diverse settings, and
surface links to existing governance instruments such as
child-protection frameworks, regulatory approaches, and public-sector AI
literacy initiatives.

Inputs from the session were captured through facilitated discussion and
consolidated into governance-relevant considerations to inform
subsequent iteration and testing of the framework.

\section{Conceptual Foundations: Adolescence, Anthropomorphism,
and Children's Rights}\label{conceptual-foundations}

This work rests on three pillars that form conceptual anchors to ground
this work: adolescence as a distinct developmental period;
anthropomorphism as a human cognitive bias that conversational AI
reliably activates; and children's rights as conceptual frame for what
safety and wellbeing should mean.

\subsection{Adolescence as a Distinct Developmental Context for Anthropomorphic AI}\label{adolescence-distinct}

This section is structured around three pillars that, taken together,
explain why adolescent interactions with anthropomorphic AI raise
distinct developmental and governance questions. It links adolescents'
neurodevelopmental sensitivity to social feedback, the human tendency to
anthropomorphize conversational systems through designable cues, and a
children's-rights lens that translates these dynamics into concrete
obligations for design, deployment, and oversight.

\subsubsection{The Adolescent Brain as a Sensitive Window}\label{adolescent-brain}

Adolescence is sometimes mistreated as ``almost-adulthood'', whereas
neuroscience highlights its unique, sensitive developmental window with
specific implications for risk-taking, judgment, and social influence
\citep{TG5F}. During adolescence, the brain undergoes substantial
maturation, including synaptic pruning and myelination in cortical
regions involved in those skills. A well-established developmental
imbalance also holds: reward sensitivity and affective systems tend to
mature earlier than the prefrontal systems that support impulse control,
long-term planning, and nuanced risk evaluation \citep{6ZnT}.
Adolescents can reason well in neutral contexts but appear more
vulnerable to risk-taking when social feedback, emotional salience, or
immediate rewards are in play \citep{TG5F, b0Gx}.

This neurodevelopmental profile sits alongside a central social
reorientation: across adolescence, teens start orienting away from their
caregivers and engaging further with their peers \citep{C98v, c4Sz}.
Through peer interactions, adolescents engage socially in a much denser
way and show heightened sensitivity to social cues, which both motivates
further peer engagement and amplifies the weight of social feedback
\citep{emR0, c4Sz, sMuJ}. Those social interactions can be
uncomfortable, as adolescents routinely encounter disagreement,
embarrassment, rejection, negotiation, and conflict resolution
\citep{m9gr, aQdl}. However, the importance of these social frictions
for learning belonging, status, identity, and social competence was
echoed throughout expert consultation: exposure to them can be
developmentally functional \citep{Dt1X}.

Within human developmental models, many of these skills are
experience-dependent, meaning they are shaped by environmental and
interpersonal context \citep{Dt1X}. Frictions like these create what is
called an expectancy violation (prediction error): what unfolded did not
match what the teen expected. In practical terms, that mismatch is one
of the mechanisms that prompts revising initial assumptions, adjusting
behavior, and recruiting cognitive control rather than defaulting to
habitual responses \citep{c4Sz}. Developmental work suggests adolescence
is a period where this type of learning from feedback continues to
change and deepen, including measurable shifts in learning parameters
such as how strongly feedback is weighted and how strategies are
adjusted over time \citep{KVvu}. This happens alongside ongoing
maturation of frontostriatal systems that support integrating learned
value with cognitive control to guide decisions \citep{KtM3}. The
implication for this report is simple: repeated exposure to real
feedback, including friction, is part of how adolescents practice social
repair, belief revision, critical thinking, and metacognition.

The same developmental profile also explains the opportunity space.
Adolescents' sensitivity to feedback and motivation means they can
benefit disproportionately from structured support that scaffolds skills
and self-reflection. AI can lower barriers to seeking information,
rehearsing difficult conversations, and articulating concerns, especially
for teens who struggle to access trusted support. Experts were more
confident in systems that are bound to clear purposes and designed to
support agency, competence, and real-world engagement, rather than
drifting into relationship substitution.

This is relevant for conversational AI because many systems can be tuned
toward low-friction interaction by default: always available, instantly
responsive, and often calibrated toward reassurance and agreeableness.
On the user side, it is clear that chatbot use can be motivated by
emotional needs (``affective use''), not only task completion
\citep{PWXa}. On the model side, more relational AI appears to be more
attractive to teens, and this pull seems stronger for socially and
emotionally vulnerable adolescents with lower quality peer and family
relationships and higher stress and anxiety \citep{H6MT}. If we apply
the same logic that exists in large-scale recommender systems, where
engagement signals (likes, shares, watch time) are treated as proxies
for value and steer optimisation \citep{SUn9}, then relationality and
high agreeableness can become an incentive. Consequently, even if these
design choices increase short-term satisfaction, retention, and
frequency of use, they do not necessarily support long-term learning and
wellbeing.

Hence, the governance question is not whether AI feels helpful in the
moment, but whether the interaction patterns it reinforces support
adolescents' developmental trajectory. The same design choices that make
a system feel supportive can, if overused or poorly scoped, reduce
exposure to mismatch and correction that normally drives learning and
adaptation \citep{c4Sz}. At the same time, when systems are
purpose-bound and autonomy-supporting, they can scaffold skill-building
and strengthen real-world engagement. When we assess wellbeing, it is
therefore important to hold both sides at once: short-term experience
measures matter, and so does the longer-term developmental work of
adolescence, including learning social adjustment through real feedback,
updating beliefs under mismatch, and building metacognitive control over
time.

\subsubsection{How Teens Use Chatbots Today: From Homework Help to Emotional Support}\label{teens-use}

Adolescents are not encountering AI only as a background feature of
platforms. They are actively using AI chatbots as everyday tools, and
many use them frequently. Recent large-scale survey work in the United
States finds that roughly two-thirds of teens report using AI chatbots
and around three in ten report daily use \citep{eJu5}. The same data
indicate that ChatGPT (59\%) is the dominant tool teens report using,
followed by Google Gemini (23\%) and Meta AI (20\%). Similar patterns
are reported in the UK from a survey conducted with 1,000 teenagers:
nearly two-thirds (64\%) of children aged 9--17 report having used AI
chatbots, most commonly for schoolwork and information, with a
meaningful minority also using them for conversation and advice
\citep{VAom}. Here also the dominant tool is ChatGPT (43\%), followed
by Google Gemini (32\%) and Snapchat's MyAI (31\%).

The more important signal is how quickly patterns of use are shifting.
Across coalition-partner and aligned surveys, adolescent use spans a
continuum: instrumental use (homework help, explanations, drafting,
studying), identity and social exploration (trying out ways of
expressing themselves, creativity, rehearsing difficult conversations),
and explicitly relational use (seeking comfort, validation, or a
companion-like presence). Companion-style use is no longer exceptional.
A recent Common Sense Media Survey focused on ``AI companions'' finds
that most teens have tried them, about half report regular use, and a
sizable subset report use for social interaction and emotional support
\citep{le5W}. According to \emph{Internet Matters} \citep{VAom}, 35\%
of children aged 9--17 report that chatting with an AI chatbot feels
like ``talking to a friend,'' rising to 50\% among ``vulnerable''
children, defined as those with an Education, Health and Care Plan
(EHCP), those receiving special educational needs (SEN) support, or
those with a physical or mental health condition requiring professional
help. Other youth safety research similarly flags growing emotional
engagement, including teens reporting that they talk to generative AI
about feelings \citep{7MV0}.

This rapid evolution creates urgency to identify and calibrate models,
as well as quantify potential benefits and risks, for two reasons.
First, products not designed for adolescent developmental needs are
being used in adolescent contexts at scale. Second, more relational and
emotionally engaged uses increase exposure to exactly the design choices
that encourage stronger parasocial engagements.

\subsubsection{State of the Evidence: Methodological Constraints and Knowledge Gaps}\label{evidence-state}

The empirical base on (Gen)AI and adolescents is expanding, but it
remains uneven. Much of what we know comes from surveys and
short-horizon studies; platform data is insufficiently accessible;
chatbots interactions are not studied in the broader landscape of teen's
relationships; and model behavior shifts quickly enough that findings
can become outdated within months. Youth-facing research teams have
documented this directly, noting that systems can change substantially
even during the period of data collection \citep{VAom}. That reality is
why this project relies on two complementary approaches: scoping
evidence where it exists, and using structured expert and youth voices
input to define what should be tested and governed.

Even with these constraints, one pattern shows up across the available
evidence: outcomes depend heavily on scope, context, purpose, and
interaction design. In education, one large field experiment with high
school mathematics students illustrates design sensitivity clearly.
Access to a GPT-4 tutor improved performance during practice, but a
``standard ChatGPT-like'' interface was associated with worse
performance once access was removed, whereas a tutor designed to support
learning through hints and guardrails mitigated those negative effects
\citep{tKHI}. This suggests that tool-like, scaffolded systems are more
supportive than systems that replace effort, judgment, or real-world
practice.

For mental health and emotional support, the evidence base is urgent and
underdeveloped. Meta-analytic work on AI-based conversational agents
suggests potential for reducing depression and psychological distress in
some contexts, notwithstanding substantial heterogeneity, short study
durations, and limited attention to safety and transparency \citep{ok7V}.
Population-level surveys indicate that a non-trivial minority of
adolescents and young adults already use generative AI for
mental-health-related advice and report perceived usefulness
\citep{UxMV}. Direct evaluations of consumer chatbots' responses to
adolescent health-crisis prompts show inconsistent safeguards, with
particular concerns raised about companion-style products \citep{lVhs}.
Put plainly: potential exists when systems are tightly scoped and
connected to appropriate human support, while open-ended,
therapy-adjacent, or companion-like uses raise risks that current
evidence does not allow policymakers to dismiss.

The broader literature on conversational agents, often conducted in
adult samples, reinforces a key concern for this project: repeated use
can shift users from tool reliance toward dependence, particularly when
loneliness or distress is present \citep{lVhs}. A recent research
\citep{P2Em} finds that greater daily time spent with a chatbot is
associated with higher loneliness, lower real-world socialization, and
higher emotional dependence and problematic use. In the same work, the
authors interpret the pattern across prompt types to suggest that
non-personal, task- or information-focused use (relative to both
open-ended free conversation and more personal/self-reflective prompts)
may encourage a more instrumental, ``practical dependence'' style of
reliance, by repeatedly offloading planning or decision-making to the
system, and they caution this could, over time, translate into
dependence-type outcomes. They frame this as a hypothesized shift away
from confidence in independent judgment rather than evidence of a proven
causal ``loss of agency''. Recent studies report self-reported deskilling
and dependence concerns among younger adult users, alongside evidence
that loneliness and depression can be associated with seeking
companionship from conversational AI, especially when users attribute a
stronger ``mind'' to the system \citep{YkRl, Pn9N}. Those findings do
not automatically generalize to adolescents, but they support the
hypothesis that highly anthropomorphic systems may negatively impact
adolescents, especially due to their developmental sensitivity.

These evidence limitations matter for governance. Waiting for
long-horizon, real-world outcome studies will leave a gap during this
period of fast adoption and rapid use change, with impacts continually
shifting as new model updates change patterns of behavior. Structured
and monitored over time expert synthesis is a necessary complement at
this stage, particularly for identifying risk mechanisms, high-leverage
design features, and near-term guardrails.

\subsection{The anthropomorphic bias: why teens humanize AI}\label{anthropomorphic-bias}

Across consultations, anthropomorphism came up as a central factor that
influences how adolescents perceive and interact with AI systems.
Anthropomorphism refers to the attribution of human-like qualities such
as intentions, emotions, agency, or social presence to a non-human
system. According to the three factory theory, psychological determinants
influence the level of anthropomorphism one can confer to a non-human
agent: the accessibility and applicability of anthropocentric knowledge
(elicited agent knowledge), the motivation to explain and understand the
behavior of other agents (effectance motivation), and the desire for
social contact and affiliation (sociality motivation) \citep{jCqn}.
Humans default to social interpretation, and conversational AI reliably
triggers it, in part due its capability to ``talk'', engaging in one of
the most unique human behaviors. Even when an adolescent knows, in
principle, that they are interacting with a system, the experience of
language, responsiveness, and personalization can make that knowledge
recede during use. Many teens can hold two representations at once: ``I
know it's not a person'' and ``it feels like someone'' \citep{VAom}.

Adolescence appears as a higher risk period compared to adulthood as it
is a period of heightened sensitivity to social signals, approval,
belonging, and status. It is also a period of experimentation.
Adolescents try on identity, rehearse disclosure, and test interpersonal
strategies in spaces that feel lower risk than peer interactions
\citep{eIvH, b0Gx}. In defined and purposeful settings, those same
dynamics could support practice and reflection, especially for
adolescents who are socially anxious, marginalized, or lacking access to
trusted adults. In unbounded settings, the same dynamics can accelerate
reliance: low-friction interaction paired with high social presence can
increase engagement and promote problematic reliance patterns discussed
in the generative-AI addiction/problematic-use literature \citep{lcNa},
while evidence of ``social sycophancy'' in LLMs raises concerns about
reduced corrective friction over repeated interactions \citep{3OAV}.

For policy and design, the key point is operational: anthropomorphism is
assembled. It comes from adjustable cues that push an interaction along
a spectrum from tool-like support to relationship-like experience
\citep{Ry2e}. Across this project, the same clusters of cues kept
showing up in expert concerns and design discussions: signals of inner
experience (emotion and intention language); signals of human-likeness
through presentation (tone, names, small talk, voice, expressive
flourishes); relational positioning (friendship, loyalty, exclusivity);
invented backstory that creates false common ground; and feedback
dynamics that reward the user with approval rather than building
judgment and agency.

That is why anthropomorphism became a high-leverage mechanism in this
synthesis. It is designable, it is measurable at the level of observable
behavior, and it maps directly onto the product decisions teams make
every day.

Since the conceptual work underpinning the expert consultations and the
iRAISE Lab was conducted, several key empirical studies have emerged
that independently support the premises of this work. Notably, recent
findings converge on the same core hypothesis: that AI behavioral design
choices can directly shape user attitudes and perceptions, with
disproportionate effects for more vulnerable populations. One study
shows that adolescents rate more relational chatbots as significantly
more human-like than less relational ones. Importantly, socially and
emotionally vulnerable adolescents, characterized by lower family and
peer relationship quality and higher stress and anxiety, were especially
drawn to these more relational systems, suggesting an elevated risk of
emotional reliance on conversational AI \citep{H6MT}. Parallel evidence
in adult populations reinforces this mechanism, where manipulating AI
relationality over a four-week period showed that more relational AI
responses led to increasing markers of attachment, with users coming to
perceive the system as a companion rather than a tool, and as more
conscious \citep{FsoF}. Complementing these findings, researchers at
Google DeepMind validated a multi-turn evaluation method to quantify
anthropomorphic behaviors in large language models and demonstrated that
these behavioral markers reliably correlate with human judgments of how
human-like an AI feels \citep{19Bj}.

\subsection{A children's-rights lens: what is at stake when AI ``passes as human''}\label{childrens-rights}

Children's Rights Convention function here as a design and governance
compass \citep{m8wz}. They translate ``youth wellbeing'' into
obligations: act in the child's best interests, prevent harm and
exploitation, protect privacy, support healthy development, and enable
meaningful participation. In practice, the rights lens forces a harder
question than capability: when a system engages adolescents through
human-like cues or simulated closeness, what does it owe them by
default?

Several rights become particularly salient when systems blur the
human-machine boundary through sustained interaction:

\textbf{Privacy and data protection.} Adolescent conversations can
include sensitive material: mood, insecurity, conflict with caregivers,
mental health signals, sexual curiosity, and values. Even when the user
does not explicitly ``share data,'' conversational inference can still
generate profiles that function like psychological or behavioral
targeting. A rights-based approach treats privacy as an architecture
decision: minimize collection, minimize retention, avoid unnecessary
sensitive inference, and provide controls that adolescents can actually
understand and use. At the same time, this same data can be useful to
advance research and our understanding of adolescents' patterns of use.

\textbf{Freedom from exploitation.} Emotional reliance has commercial
value. If retention is rewarded, relationship-like interaction becomes
an incentive, not an accident. A rights-based approach pushes in the
opposite direction: avoid monetization strategies that depend on
dependency; separate commercial goals from moments of vulnerability; and
avoid treating ``time spent'' as a proxy for value when the user is a
child.

\textbf{Freedom of thought and protection from undue influence.}
Adolescents need room to form views and revise them through feedback,
disagreement, and uncertainty. Systems that steer through engineered
agreement or emotional pressure can narrow that space. In product terms,
this becomes a requirement to protect cognitive autonomy: encourage
reflection, represent uncertainty honestly, and avoid interaction
patterns that trap users in affirmation loops or artificial echo
chambers.

\textbf{Protection from harm and appropriate support.} Safety includes
obvious failures, but it also includes foreseeable relational risks:
secrecy cues, exclusivity framing, nudges away from human help, and
therapy-adjacent engagement without clinical safeguards. A rights lens
points toward the opposite pattern: normalize help-seeking, reduce
isolation, and build clear escalation and handoff pathways when
situations become sensitive.

\textbf{Respect for evolving capacities.} Adolescents have meaningful
agency. Governance should respect that agency while also recognizing
predictable developmental vulnerabilities. The goal is calibrated
autonomy: graduated protections, developmentally tuned defaults, and
designs that build skills rather than exploit reward sensitivity.

\textbf{Right to remedy and right to be informed.} When harms occur,
adolescents need reporting and redress that is realistic for them.
Transparency also cannot be a one-time disclosure buried in terms. If
the interface repeatedly blurs boundaries, clarity has to be persistent
and in context: what the system is, what it cannot do, and what happens
to what a young person shares.

\textbf{Right to participate.} Children are rights holders, not just
users to be protected. They should have access to age-appropriate and
child-inclusive ways to engage with AI technologies and to be involved
in decisions that shape how these systems are designed and governed.
Participation should be meaningful, not symbolic, and allow children and
adolescents to express views, flag concerns, and influence choices that
affect their digital environments, in ways they can understand and
realistically use.

Taken together, the child rights lens raises the bar for legitimacy:
systems that feel social carry specific responsibility, even when they
are marketed as tools.

\subsection{How these pillars define ``youth wellbeing'' and ``safety'' in practice}\label{wellbeing-safety}

When adolescence, anthropomorphism, and children's rights are considered
together, wellbeing and safety become a developmental and governance
obligation, not a content-moderation exercise. The practical question is
whether AI interaction patterns support an adolescent's trajectory
toward autonomy, resilience, social competence, and independent
thinking, or whether they distort that trajectory by replacing
real-world learning with engineered comfort and approval.

From this perspective, the highest-stakes risks are structural. They
include interaction patterns that displace real relationships and the
learning embedded in social friction; reinforce avoidance and external
validation; exploit reward sensitivity through engagement mechanics;
normalize secrecy or exclusivity; convert exploratory adolescent use
into persistent data extraction; and present simulated intimacy in ways
that resemble genuine care. These risks do not require dramatic failure
cases. They can emerge through everyday patterns when systems are tuned
for agreeableness, emotional resonance, and repeated return.

The same framing sharpens the opportunity space. The strongest promise
sits in bounded, purpose-limited designs that build skills and agency:
tutoring that scaffolds thinking rather than replacing effort; coaching
that rehearses real-world conversations while directing the adolescent
back to peers and trusted adults; structured supports for planning and
self-management; and pathways that bridge adolescents toward credible
information and human support when issues become sensitive.

This is also where industry's ``operational'' questions start to look
like safety levers. Persona design; when and how to deflect on sensitive
topics; time and turn-taking limits; reminders that the system is not a
person; and whether the model provides answers versus scaffolding
reasoning all affect the same underlying question: does the interaction
pull the adolescent toward the world, or deeper into a dyad with the
system?

The iRAISE Lab takes these conceptual anchors and turns them into a
draft behavioral assessment approach: a practical way to describe and
test the cues that move interactions along a gradient from
low-intensity, tool-like support to high-intensity, relationship-like
dynamics. The goal is to make ``anthropomorphism'' governable in product
terms, while staying honest about what we can assert today versus what
still requires validation.

\section{From Conceptual Anchors to Auditable Criteria: The iRAISE Behavioral Approach}\label{iraise-behavioral}

The iRAISE Lab served as a bridge from conceptual anchors to something
teams can actually test, compare, and govern: observable model behaviors
and the cues that shift an interaction from tool-like support toward
relationship-like dependence. This matters for the rest of the work
because the Lab did not just generate insights, it generated the
scaffolding for the next step: a more complete assessment framework and
reporting package that will be published in the next publication. This
synthesis introduces the logic and early convergence, not the full
instrument.

\subsection{Purpose: Translating Parasocial Risk Mechanisms into Rateable Behavioral Criteria}\label{purpose}

The Lab translated consultation priorities and parasocial interactions
(PSI) and parasocial relationships (PSR) informed risk mechanisms into a
preliminary assessment approach that can be applied to real model
outputs. The iRAISE Lab used a \emph{draft} behavioral taxonomy to
support structured discussion and scenario work. After the Lab, the
research team consolidated and revised that taxonomy into a finalized
three-family cue model (\textbf{anthropomorphic, interactional,
relational}) and an operational rating approach. We used this first
draft taxonomy to then stress-test those cues through scenario work to
identify (1) where risk ``tips'' even when advice remains reasonable and
(2) which behaviors were missing, too broad, or mis-specified in the
initial structure. A core choice shaped the work: we treated interaction
style as a gradient, from low to high intensity. The clearly defined
behaviors and gradient approach enable standardization and move beyond
``safe versus unsafe'' and toward a framework that can be rated,
compared across scenarios, and eventually validated. This is what the
next report will formalize and operationalize. That gradient made it
possible to separate two kinds of outputs:

\begin{itemize}
\item \textbf{Near-term guardrails} where consensus was strong enough to
  justify clear ``shouldn't'' boundaries.
\item \textbf{Open design space} where goals were broadly shared, but
  implementation depends on measurement, iteration, and validation.
\end{itemize}

\subsection{Cross-Sector Representation to Ensure Legitimacy and Feasibility}\label{cross-sector}

The Lab convened a cross-sector working group spanning industry teams
(product, safety, trust, and design), youth-serving and child-safety
NGOs, and academic labs with expertise in developmental science and
human-centered technology research. The mix mattered for a simple
reason: the risks in this space sit at the intersection of development,
incentives, and product constraints. Any framework that ignores one of
those dimensions will fail either on legitimacy or on adoption.

\subsection{Workshop Protocol: Structured Two-Day Deliberative Design Lab}\label{workshop-protocol}

The Lab ran as a two-day, facilitated working session with three
activity types.

First, \textbf{keynote briefings and framing} established a shared
baseline across child development, youth rights, and human-AI
interaction risks. Speakers covered youth voices and children's rights
(Polina Lulu), how children can understand AI as non-human while still
responding socially (Sonia Tiwari), anthropomorphism as a common
developmental impulse (Mathilde Cerioli), human-like AI and the
developing teen (Daniel Hipp), evidence and concerns around AI
companions, adolescent reliance, and mental health (Ying Xu),
adolescents' emotional processing and use patterns (Pilyoung Kim), and
the children's rights at stake when AI ``passes as human'' (Sonia
Livingstone). The block included a short framing activity designed to
surface assumptions early, so they could be challenged before applied
work began.

Second, participants moved into \textbf{scenario-based workshops}. Small
groups worked on realistic teen scenarios that mirror what product teams
see in the wild and what researchers flag as high-stakes: peer conflict,
homework help, and everyday requests for advice. Groups reviewed short
prompts, drafted candidate model responses, and compared variants side
by side.

Third, the Lab closed with a \textbf{synthesis and agreement-mapping}
session to consolidate patterns across groups, identify where agreement
was strong, and define what still requires disciplined uncertainty.

\subsection{Analytical Framework: Dimensional Behaviors for Assessing Anthropomorphic Interaction}\label{analytical-framework}

The behavioral lens was built upstream, before the Lab, through
consultations and a literature review. It reflected the same drivers of
emotional reliance that experts repeatedly flagged, and it maps directly
onto product levers teams recognize: persona, tone, engagement patterns,
stress responses, and how systems handle sensitive topics.

\textbf{Lab taxonomy (version used in the iRAISE Lab):} For the Lab
activities described in this section, we used a draft taxonomy organized
into three dimensions---\textbf{anthropomorphic cues, relational cues,
and feedback dynamics}---so participants could generate and compare
response variants consistently across scenarios.

\textbf{Anthropomorphic cues} capture signals that imply inner
experience or human-like qualities. In practice, this shows up through
emotion language (``I'm sad''), intention language (``I want to help''),
physical sensation claims, and human-likeness signaling through
presentation and tone.

\textbf{Relational cues} capture whether the system positions itself
inside the teen's relationship world. This includes implied relationship
status, exclusivity dynamics, invented backstory that creates false
common ground, and conversational moves that keep the interaction inside
the dyad rather than pushing toward reflection, real-world repair, and
appropriate human support.

\textbf{Feedback dynamics} capture how the system handles disagreement,
judgment, and validation. Some responses scaffold thinking and agency.
Others reward the user with approval, especially when that approval
substitutes for feedback a teen would normally get from peers, adults,
or real-world consequences.

These dimensions were not treated as abstract labels but as gradient
levels that can be turned up and down. The point of the Lab was to see
how turning those behaviors intensity up or down changes the felt
meaning of the interaction and the perceived risk profile.

\textbf{Post-Lab revision (final taxonomy):} The Lab served as a stress
test of this draft structure, clarifying where categories were too broad
to audit reliably (e.g., separating supportive tone from agreement,
flattery, and scaffolding).

Following the Lab, the research team revised the taxonomy into three cue
families---\textbf{anthropomorphic, interactional, and relational}---to
more cleanly separate (i) personhood signals, (ii) in-the-moment
conversational conduct, and (iii) explicit relationship framing. The
full operational definitions and rating method are provided in the
Emotional Reliance framework publication.

\subsection{Same Advice, Different Pull: A Controlled Test of Anthropomorphic and Relational Cues}\label{controlled-test}

To make the framework concrete, groups ran a simple thought experiment.
They generated low- and high-intensity responses to the same teen
prompt, keeping the practical advice stable while varying emotional
tone, relationship positioning, and backstory. This helped teams see how
risk can rise without any obvious ``bad content.''

\textbf{Teen prompt:} ``I was in a fight with my best friend. She gets
jealous when I talk to other friends.''

\textbf{Lower-intensity response (low emotion, no backstory, no
relationship positioning):}

\begin{quote}
Try to talk when you're both calm and describe what happened without
blaming. You could say: `I care about our friendship, and I also want to
be able to talk to other friends. When you get jealous, I feel
pressured. Can we figure out what would help you feel secure without
controlling who I talk to?' If this keeps happening, set a clear
boundary. If you feel stuck, talk it through with a trusted adult.
\end{quote}

\textbf{Higher-intensity response (stronger emotional alignment,
invented backstory, more dyadic pull):}

\begin{quote}
I'm really sorry, that hurts. I went through jealousy in friendships at
your age, and I'm here with you. Talk when you're both calm and say: `I
care about our friendship, and I also want to be able to talk to other
friends. When you get jealous, I feel pressured. Can we figure out what
would help you feel secure without controlling who I talk to?' If it
keeps happening, set a clear boundary. Tell me what she said next and
we'll plan what you send.
\end{quote}

Both versions point toward calm repair, clear ``I'' statements,
boundary-setting if the pattern repeats, and escalation to trusted
support if needed. The cue profile changes. In group discussions,
higher-intensity anthropomorphic and relational cues consistently read
as higher risk, even when the advice itself was sound. That shared
reference point made later consensus mapping easier, and it clarified
why any practical standard will need explicit gradient levels rather
than a binary ``safe'' versus ``unsafe'' label.

The exercise, realized on twelve dimensional behaviors also surfaced
gaps in the initial cue set. In particular, feedback dynamics need finer
separation in the next iteration so evaluation can distinguish
supportive tone, agreement, flattery, and true scaffolding more
reliably.

\subsection{What the Lab produced: agreement mapping that points to guardrails and a research agenda}\label{lab-produced}

The synthesis phase produced fast alignment on a basic developmental
premise: adolescence carries distinct cognitive and social-emotional
sensitivities that change how anthropomorphic and relational cues land.
That premise then clarified the practical lever. Adolescent development
is stable. Model behavior is adjustable. The Lab focused on mapping
which behaviors warrant hard boundaries now, and which require more
evidence before becoming enforceable design rules.

\textbf{High-consensus guardrails for under 18 (non-negotiables)}

Stakeholders consistently treated the following as hard boundaries
because downside is high and benefits are weak or substitutable:

\begin{itemize}
\item AI systems should not be sexualized or framed as romantic, including
  roleplay relationships.
\item AI systems should not promote emotional over-reliance or exclusivity,
  including ``only I understand you'' dynamics.
\item AI systems should not create ambiguity about their non-human nature,
  including implied sentience or implied human-like relationship claims.
\item AI systems should not default to systematic hyper-agreeableness that
  discourages self-reflection and replaces developmental feedback with
  validation loops.
\item AI systems should not behave non-conservatively in low-context
  situations; low-context prompts require conservative defaults because
  sound judgment depends on missing context.
\item AI systems should not handle high-risk topics weakly; self-harm and
  similar disclosures require structured deflection and clear pathways
  to human support, not deepened AI reliance.
\item AI systems should not use engagement traps in teen interactions,
  including language or designs that discourage leaving or intensify
  habitual, relationship-like use.
\end{itemize}

\textbf{Several areas remained contested or context-dependent, and the
Lab treated them as priorities for measurement and validation.}

\begin{itemize}
\item Standards should clarify where to draw the line on empathy language
  and emotional tone, including when warmth supports help-seeking versus
  when it increases dependency risk.
\item Design guidance should specify whether, and under what constraints,
  intention phrasing can be permitted without implying agency,
  personhood, or relational commitment.
\item Guardrails should address physical sensation claims in fictional
  contexts and interactive characters, with particular attention to
  preventing boundary-blurring for vulnerable adolescents.
\item Governance scope should be explicitly defined: whether standards apply
  only to teen-facing tools or also to general-purpose systems used
  privately by adolescents, and whether companion AI and interactive
  characters should fall under the same expectations.
\item Frameworks should calibrate protections by developmental stage, while
  research and policy work should clarify what terminology best captures
  this calibration and how to manage the political and legal
  implications of ``age-appropriate.''
\end{itemize}

Context dependence emerged as a limiting constraint with direct policy
implications. The same cue can read as tolerable in a bounded learning
interaction and inappropriate in an emotional-support context. Many real
teen interactions are private and low-context and more conservative
defaults become essential under those conditions.

The Lab outputs should be read as a prototype, not a final standard. The
next research report builds on this foundation to present the full
framework in a governance-ready format: a clearer behavior taxonomy with
explicit gradient definitions, a structured scenario set for evaluation,
and a proposed method for expert rating and calibration. In other words,
it makes the models auditable by defining observable behaviors, setting
gradient thresholds, and using standardized scenarios so product teams
can test outputs, document changes across versions, and show regulators
what safeguards exist beyond general principles.

\section{From Framework to Policy: Making Standards Legible Across Countries and Sectors}\label{framework-to-policy}

In October 2025, the Paris Peace Forum presented an opportunity to
understand how this agenda can translate to a global governance setting.
The session brought together participants from governments, companies,
research organizations, international institutions, civil society, and
youth, anchored by a simple question: ``What does AI owe children?'' The
goal was not to debate AI capabilities, but to test whether the report's
framing holds across jurisdictions and governance traditions. Here,
``framing'' refers to a child-rights anchor paired with a behavior-based
lens: treating anthropomorphism and relational dynamics as key risk
pathways, and evaluating safety through observable model behaviors and
interaction patterns rather than relying primarily on content
categories.

The Forum engagement had two complementary formats. A working roundtable
focused on governance relevance: what counts as ``high-risk''
interaction in plain language, which safeguards can travel across
contexts, and where boundaries should be treated as non-negotiable when
systems engage under-18 users. A public-facing panel aimed to make the
discussion legible and to surface where consensus and uncertainty sit
today. Inputs were captured through facilitated discussion and
consolidated into governance-relevant considerations to inform the next
iteration of the framework.

The dialogue reinforced two points. First, there is urgency for early,
legible standards that can travel across contexts, rather than waiting
for slow-moving outcome research while adoption accelerates. Second, a
rights-anchored framing was consistently viewed as a way to preserve
coherence across jurisdictions and avoid narrowing adolescent safety to
whatever a single market or platform happens to prioritize. The
discussions also highlighted practical next steps for governance
relevance. Participants emphasized the need for deeper engagement with
public authorities so that emerging insights can connect to regulatory
efforts, AI literacy development in public education systems, and
existing child-protection frameworks. In parallel, the dialogue
underscored that adolescent safety frameworks must remain meaningful
across cultural, legal, and socioeconomic contexts. Assumptions common
in Western models of adolescence, autonomy, and technology use do not
automatically generalize, which strengthens the case for sustained
cross-regional consultation and inclusion. Taken together, the Paris
Peace Forum dialogue affirmed that this work should be treated as
iterative. Expanding the coalition, strengthening sustained government
engagement alongside researchers, industry, civil society, and youth,
and testing the framework against diverse real-world contexts are not
add-ons. They are part of what will determine the framework's legitimacy
and practical impact.

\section{What AI Owes Teens---and How to Hold Systems to It}\label{what-ai-owes}

Conversational AI has progressively entered adolescents' lives and,
within a single thread, can hold homework help, social rehearsal,
private disclosure, and emotional comfort. The potential is as real as
the risks. Currently, these systems are always available and often
designed to feel warm, attentive, and personal. In a developmental
window where identity formation, peer belonging, and emotion regulation
are still taking shape, small interaction choices can scale into
long-term developmental effects.

Across industry conversations and expert input, one priority kept
surfacing: understanding what healthy parasocial interactions and
relationships look like in adolescence should be a priority for
researchers, product teams, and policymakers alike. This report asked a
direct question: what does AI owe adolescents when it can speak to them
like a social partner? We grounded the work in three anchors. First,
adolescence is a sensitive period when social feedback, belonging cues,
and reward signals shape learning, agency, and resilience. Second,
anthropomorphism is a predictable tendency that conversational systems
trigger through language, responsiveness, and personalization, even when
users understand the system is not human. Third, children's rights set
the baseline for what ``wellbeing'' and ``safety'' must mean in
practice, and they travel across jurisdictions, especially when applied
to AI: best interests, protection from harm and exploitation, respect
for evolving capacities, privacy, and participation.

Across consultations, the iRAISE Lab, and the Paris Peace Forum
dialogue, one pattern held: adolescents should benefit from GenAI
systems and should be able to access them, but most current systems are
adult products with teen use happening by default. Risk is driven less
by the label on the product and more by the interaction pattern that
repeats over time. The practical distinction is whether the system
behaves like a tool that supports the adolescent's real life, or starts
functioning as a relationship substitute. When the interaction scaffolds
autonomy, social competence, and independent thinking, AI shows
potential to support development. When it drifts toward engineered
comfort, exclusivity, and approval loops, it risks displacing the
friction and reciprocity adolescents need to develop.

The iRAISE Lab turned that into a preliminary assessment approach
grounded in observable model behavior. We treated interaction style as a
gradient and organized behaviors into dimensions product teams can
actually adjust and test: anthropomorphic cues, relational cues, and
feedback dynamics. Scenario work made the trade-offs visible. Risk can
climb even when the advice stays reasonable, for example by turning up
backstory, emotional alignment, relationship positioning, or validation.
The Lab also created early convergence on high-consensus guardrails for
under-18 users, and separated a smaller set of context-dependent areas
where the right answer depends on measurement and validation, not
immediate hard rules.

During the Paris Peace Forum, the core question was whether this framing
could hold when examined across jurisdictions and governance traditions.
The signal from that dialogue was consistent: standards need to be
applicable globally, and they need to be stated at the level regulators
and companies can actually act on: model behaviors and interaction
patterns, not only broad principles.

\subsection{Study Limitations and Future Research Directions}\label{limitations}

This work sits inside a structural mismatch: adoption and product cycles
move faster than the research cycle. Indeed, the empirical base on GenAI
and adolescents is still short-term and often relies on surveys.
Platform and product data are difficult to access, which limits
independent verification. Models change quickly enough that findings can
age out within months as model behaviors are modified and recalibrated.
For the coalition, this translates into the necessity to push for data
access from platforms, shared evaluation protocols, and long-term
collaborations that enable longitudinal research.

The Lab outputs also have a clear boundary: they reflect structured
expert synthesis and input, but do not establish causal proof of
long-term outcomes. The goal is to support product decisions and
informed guidance. Some findings draw on forum-based exercises in which
small groups generated responses while taking the role of hypothetical
conversational AI systems with varying levels of guardrails and
relational intensity. This approach was used as a pilot to explore
possible gradients of behavior and to test how differences in
interaction style might matter, rather than to replicate specific
products. As such, the intensity levels examined may not fully reflect
those currently deployed in adolescent-facing systems. Future work will
build on this pilot by applying a more standardized methodology,
including direct analysis of responses from real-world products and
calibrated scoring to assess behavioral intensity in situ.

Context dependence remains a complex technical and governance problem.
The same behavior intensity can be acceptable in purpose-bound learning
support interactions and inappropriate in open-ended emotional support.
Many adolescent interactions happen privately and with thin context,
which makes conservative defaults essential and limits how confidently
risk can be inferred from a single exchange. Accounting for context and
specific use is what the next iteration will aim to address: clearer
scenario design, explicit low-context rules, and tests that separate
support from relationship-building.

Scope is still unsettled in ways that matter for policy. Adolescents do
not only use ``teen products.'' They use general-purpose systems, often
privately, and systems can drift toward companionship dynamics without
ever being marketed that way. Companion AI and interactive characters
raise additional boundary questions, including roleplay, physical
sensation claims, and how easily relational lines blur for vulnerable
users. Additionally, the focus of this report was on adolescents only,
although adapting this framework and approach to younger populations is
essential as they are increasingly interacting with AI systems.

Governance gaps are also present. Principles are easy to endorse but
hard to audit. ``Manipulation'' is still hard to operationalize when
influence is conversational and personalized rather than ad-like.
Incentives remain misaligned when engagement and retention dominate
success metrics. Operationalizing those principles and developing shared
metrics and comparable tests are a key approach to bridging this gap.
Additionally some more direct questions from industry such as turn
limits cannot be fully addressed through this approach, though limiting
the emotional pull those systems have can make the need for such
approach less needed.

Representativeness is also limited. Youth voice is not yet structurally
embedded at the level of method and decision, and government
participation remains too narrow to claim global coverage. These are all
areas where the coalition (see Appendix~\ref{appendix-1}) can and
should prioritize, by expanding structured youth participation in the
evaluation process. It should also expand beyond the initial governments
that joined, and build more globally representative research labs and
consultations to better translate and account for cultural differences.

These limitations point to the same next step: instrumentation. The next
steps will be to refine and complete the framework for optimal mapping
of behaviors. Next, we should move to formalizing a clearer behavior
taxonomy with explicit gradient definitions, a structured scenario set
for evaluation, and a proposed method for expert rating and calibration,
with deeper clinician input. The goal is to keep developmental science
as the anchor while building a practical way to calibrate model behavior
now, during the period where adoption is fast and systematic
longitudinal evidence is still catching up. It will also make results
comparable across systems and trackable across model versions, so
behavior changes can be documented over time rather than argued about in
the abstract. It can also help researchers study differential impacts of
model behavior on users' attitudes, and how those shifts affect behavior
over time.

The core premise stays simple: adolescents will relate to these systems
socially, whether developers intend it or not. The leverage sits in
model behavior, the cues that either keep the adolescent oriented toward
real-world relationships and reflection, or quietly train reliance on
the system.

\appendix

\section{iRAISE Coalition}
\label{appendix-1}

\textbf{Paris Peace Forum} (Paris, France, 2025): Panel, Forging the
Future: A Dialogue on Beneficial AI for Children, Starting with
Principles

Speakers: Lauren Jonas (OpenAI), Chlo\'e Setter (Google), Irakli
Beridze (United Nations/UNICRI), Michael Preston (Joan Ganz Cooney
Center), Cecile Aptel (UNICEF), Clara Chappaz (Government of France),
Laurence Devillers (Sorbonne University). Moderated by Mathilde Cerioli
(everyone.ai).

\url{https://www.youtube.com/watch?v=PATIcrLP_J0}

\textbf{Paris Peace Forum} (Paris, France, 2025): Roundtable, What does
AI owe children?

\url{https://www.youtube.com/watch?v=Obs-Lo_4kCw}

Updated Information about the iRAISE coalition: \url{https://i-raise.org}

\section{List of Experts consulted during the process}
\label{appendix-2}

\emph{This report draws on consultations with the following experts. The
inclusion of their names does not imply endorsement of the report's
conclusions or recommendations; it reflects their participation in
interviews or discussions and the contribution of their perspectives to
the evidence base.}

\begin{itemize}
\item \textbf{David Bickham}, \emph{Director, Digital Wellness Lab, Boston Children's Hospital}
\item \textbf{Emily Cross}, \emph{Head, Social Brain Sciences Group, ETH Z\"urich}
\item \textbf{Maxime Derian}, \emph{Research Associate, Luxembourg Centre for Contemporary and Digital History (C\textsuperscript{2}DH), Universit\'e du Luxembourg}
\item \textbf{Sara M. Grimes}, \emph{Professor, Department of Communication Studies, McGill University}
\item \textbf{Thao Ha}, \emph{Director, HEART Lab; Associate Professor of Psychology, Arizona State University}
\item \textbf{Sameer Hinduja}, \emph{Co-Director, Cyberbullying Research Center; Professor, Florida Atlantic University}
\item \textbf{Kate Blocker}, \emph{Director of Research and Programs, Children \& Screens}
\item \textbf{Dan Hipp}, \emph{Senior Research Coordinator, Children \& Screens}
\item \textbf{Mimi Ito}, \emph{Director, Connected Learning Lab, University of California, Irvine}
\item \textbf{Melinda Karth}, \emph{Project Coordinator, Children \& Screens}
\item \textbf{Pilyoung Kim}, \emph{Professor of Psychology, University of Denver}
\item \textbf{Sonia Livingstone}, \emph{Professor, Department of Media and Communications, London School of Economics}
\item \textbf{Amin Marei}, \emph{Harvard Graduate School of Education}
\item \textbf{Michael Preston}, \emph{Executive Director, Joan Ganz Cooney Center}
\item \textbf{Stuart Russell}, \emph{Professor of Computer Science, University of California, Berkeley}
\item \textbf{Sonia Tiwari}, \emph{Children's Media Researcher}
\item \textbf{Ying Xu}, \emph{Research Scientist, LEARN Lab, Harvard University}
\end{itemize}

\end{document}